\documentclass[epj]{svjour}
\usepackage{graphicx}

\DeclareGraphicsExtensions{ps,eps,epsf}

\def\bea{\begin{eqnarray}}
\def\eea{\end{eqnarray}}

\begin{document}
\title{Cellular solid behaviour of liquid crystal colloids \\
1. Phase separation and morphology}
\author{V.J. Anderson\inst{1}, E.M. Terentjev\inst{1}, S.P. Meeker\inst{2},
J. Crain\inst{2} \and W.C.K. Poon\inst{2}}
\institute{Cavendish Laboratory, The University of Cambridge,
Madingley Road, Cambridge CB3 0HE, U.K. \and Department of Physics
and Astronomy, The University of Edinburgh, Mayfield Road,
Edinburgh EH9 3JZ, U.K.}
\date{\today}

\abstract{We study the phase ordering colloids suspended in a
thermotropic nematic liquid crystal below the clearing point
$T_{\rm ni}$ and the resulting aggregated structure. Small ($150\,
\hbox{nm}$) PMMA particles are dispersed in a classical liquid
crystal matrix, 5CB or MBBA. With the help of confocal microscopy
we show that small colloid particles densely aggregate on thin
interfaces surrounding large volumes of clean nematic liquid, thus
forming an open cellular structure, with the characteristic size
of $10-100 \, \mu\hbox{m}$ inversely proportional to the colloid
concentration. A simple theoretical model, based on the Landau
mean-field treatment, is developed to describe the continuous
phase separation and the mechanism of cellular structure
formation.
} 
\authorrunning{V.J. Anderson {\it et al.}}
\titlerunning{Cellular solid liquid crystal colloids: 1.
Phase separation} \maketitle
\section{Introduction}  \label{intro}
The phase separation and ordering of multi-phase systems, and
their resulting physical properties, have been the subject of much
active research over the past twenty years. Such a high interest
in this field arises from a combination of challenging,
fundamental physics (of phase equilibria and dynamics) with a
large number of viable applications. Research in liquid crystals
has also remained active for decades, for similar fundamental and
practical reasons. However, the applications of thermotropic
liquid crystals have always focussed on display technology,
somewhat overlooking their unique mechanical properties. In the
early 1990's a new field, now called ``liquid crystal colloids'',
was opened to research by an experimental and theoretical study of
suspensions of colloidal particles in a lyotropic liquid crystal
by Poulin et al. \cite{roux}. The key idea of introducing a large
amount of mobile interface into the liquid crystal, which
generates nontrivial topological constraints and singularities,
has been fruitfully explored by further experimental and
theoretical work \cite{coll95,luben,rama,interac}.

\begin{figure}[b] 
\resizebox{0.47\textwidth}{!}{\includegraphics{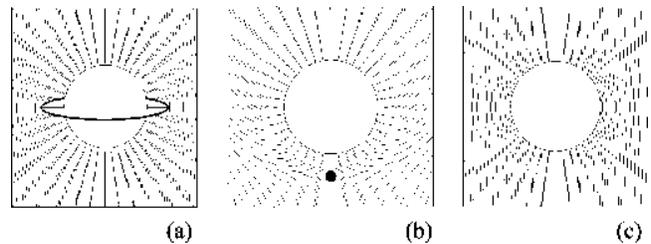}}
\caption{The calculated director field around a single spherical
particle in a uniform nematic matrix. For rigid anchoring there
are two possibilities: (a) the quadrupolar disclination ring
structure, or (b) the dipolar structure with a satellite monopole
defect. In the case of weak anchoring (c) the director field has
the quadrupolar symmetry but does not have topological
singularities.} \label{fig1}
\end{figure}

The problem studied is that of a spherical particle with nematic
director anchored on its surface, suspended in a nematic liquid
crystal matrix.  When the director is anchored rigidly, a closed
inner surface creates a topological mismatch between the director
field ${\bf n}({\bf r})$ on the particle surface and the uniform
director at large distances from it. This mismatch leads to
topological defects, i.e. regions where the liquid crystal order
and the continuity of ${\bf n}$ break down. A connected closed
surface represents a point topological charge $N=1$. Since the
overall sample has to be topologically neutral, the charge
produced by the inner surface must be balanced by an associated
opposite charge. There are two basic possibilities. The
assumption, that a spherical particle in a quadrupolar nematic
medium creates deformations that are quadrupolar too, results in
the mismatch balanced by a circular loop of disclination with
linear strength $N=-1/2$ and overall point charge $N=-1$, in the
equatorial plane of the particle: a ``Saturn ring'' \cite{ukra},
see fig.\ref{fig1}(a).
The other possibility is a dipolar configuration with a satellite
point defect, a monopole with $N=-1$ near one of the particle
poles \cite{luben}, see fig.\ref{fig1}(b). It appears that,
although both situations are possible, the dipolar structure with
satellite defect occurs more readily, e.g. in experiments
\cite{luben,poulin98} for water droplets dispersed in a
thermotropic nematic matrix. When several such droplets come to a
near-contact, the effect of chaining occurs, with satellite
defects sandwiched between inclusion particles and shared between
them, providing a strong and anisotropic interaction force.

On the other hand, when the strength of director anchoring is
rather weak, the particle introduces only a small perturbation
into the nematic matrix, see fig.\ref{fig1}(c). Albeit small, the
non-uniform distortion of {\bf n} is spreading over long distances
from the particle -- the exact solution for the angle $\theta$ of
director deviation from the uniform axis ${\bf n}_{\rm o}$ can be
obtained in terms of the multipole expansion and reads, e.g.
\cite{ukra} (in spherical polar coordinates),
\begin{equation}
\theta ({\bf r}) = \frac{1}{4} \frac{WR}{K} \left( \frac{R}{r}
\right)^3 \sin \, 2\theta \, . \label{weakT}
\end{equation}
In this expression, $R$ is the particle radius and the nematic
constants $W$ and $K$ are the anchoring energy on its surface and
the Frank elastic constant respectively \cite{degen}. Their
typical values in a thermotropic nematic liquid crystal are $K
\sim 10^{-11} \hbox{J/m}$ and $W\sim 10^{-6} \hbox{J/m}^2$, for
homeotropic director anchoring as in fig.\ref{fig1}.

The parameters in eq.~(\ref{weakT}) identify the important
dimensionless ratio which controls the outcome of the single
particle behaviour. $(WR/K)$ measures the relative effect of the
particle surface (characteristic energy $\sim WR^2$) and the bulk
director deformations (Frank energy scale $\sim KR$), and can be
large or small. When $WR/K \ll 1$ the anchoring should be
considered weak and not capable of producing large deformations in
the surrounding nematic matrix. In contrast, when $WR/K \gg 1$
rigid anchoring can be assumed and topological singularities
result. For typical parameters, particles of $R \sim 5-10 \, \mu$m
represent the border between the two topologically distinct
regimes in fig.\ref{fig1}: submicron colloid particles create
small long range director distortions of quadrupolar symmetry,
eq.(\ref{weakT}), while larger objects most frequently have a
dipolar configuration with a satellite defect.

The second physical idea relevant to our work is using the process
of phase separation in a liquid crystal phase as a means of
creating internal interfaces. The orientational symmetry breaking
and the additional curvature elastic energy in the liquid crystal
microconfined by such interfaces should have a profound effect on
the whole process of phase ordering. Both phase equilibrium and
kinetics are altered in non-trivial ways due to the underlying
frustrated liquid crystalline order.

Although the progress in the single-particle description of liquid
crystal colloids has been noticeable, little experimental work
exists in this area of colloid collective behaviour. One of the
main questions here is the macroscopic morphology of a colloid and
the change in rheological behaviour on increasing concentration of
particles, when the suspending matrix would possess a dense
network of interaction forces creating high barriers to
deformation. One may expect a glass-like freezing of motion at
sufficiently high concentrations -- an interesting possibility
that needs to be thoroughly explored.

The work of Poulin et al. \cite{poulin98} on water droplets in a
nematic matrix confirmed the ideas about topological defects and
their long-range interactions -- but they only studied the
structure around a single or few very large droplets. In contrast,
Tanaka et al. \cite{tanaka} studied phase ordering in mixture of a
nematic 5CB and a surfactant, which forms very small
nano-micelles. They reported some provocative results on a phase
they named ``transparent nematic''. One could speculate that in
this case the colloid parameter $WR/K$ is so small that the
mean-field description of nematic order would fail altogether.
Recently, Zapotocky et al. \cite{martin} examined the colloid
rheology of a cholesteric liquid crystal filled with silica
particles of size $\sim 1 \, \mu$m. They have found a variety of
rich behaviour due to topological defect-driven enhancement of
dynamic elastic properties and the formation of a weak gel with
the estimated static modulus of $G' \sim 0.01 \, \hbox{Pa}$.
However, in their work small silica particles have aggregated into
much bigger objects (flocs) that produced strong topological
forces and ended up in the nodes of a network of disclination
lines, which has been carrying the elastic function of the
resulting material.

A simpler system has been discovered recently by Meeker et al.
\cite{wilson}. A ``classical'' liquid crystal colloid has been
prepared by mixing a well-characterised thermotropic nematic
liquid crystal, 5CB, with sterically-stabilised PMMA particles,
imposing radial boundary conditions in the nematic matrix.
Mixtures of $5-10 \, \%$ particle volume fraction resulted in a
soft solid with significant storage modulus, $G' \sim 10^3 - 10^5
\hbox{Pa}$. Such a remarkable mechanical transformation requires
further investigation, which is reported in the companion paper
\cite{no2}. Bright field microscope observations by Meeker et al.
suggest that the soft solid comprises a network of particle
aggregates, formed by the exclusion of particles from emergent
nematic domains as the mixture is cooled below the
isotropic-nematic transition $T_{\rm ni}$. However, their optical
observations were limited to low particle volume fractions $(\leq
5 \, \%)$, and depended on reheating the samples back into the
isotropic phase to view the particle aggregate structure, since
direct optical study in the bulk of a highly non-uniform
birefringent system is difficult due to the strong scattering of
light: the nematic colloids are opaque below $T_{\rm ni}$. In this
paper we use a confocal microscopy technique to study directly the
structures in the bulk of the nematic liquid crystal colloid
samples, for a wide range of particle concentrations $(3 - 15 \,
\%)$.

In addition to the remarkable mechanical response, the second
aspect of interest in this system is the time required for the
formation of an aggregated state below $T_{\rm ni}$. The
characteristic time required for particle movement under the
influence of long range attraction forces has been estimated in
\cite{interac} and gives, for our example of particle size and
concentration in the limit of weak anchoring, $\tau \sim 10^{3}
\eta (K/W^2) \sim 10^3 \, \hbox{s}$ (here $\eta \sim 0.1 $~Pa.s is
the relevant viscosity coefficient of the nematic liquid crystal).
We shall see that a rigid gel is formed in a much shorter time
after the colloid mixture is brought below the nematic transition
point $T_{\rm ni}$. It appears that the phenomena observed in our
liquid crystal colloid system cannot be purely due to the
topological defects and their networks: their elastic energy is
not enough to explain the modulus and, more importantly, there
should be no topological defects at small $WR/K$.

Here we study the formation process of the resulting near-solid
liquid crystal colloid aggregates in more detail. We show that a
process of continuous phase separation of small particles takes
place in our system. This separation occurs immediately after the
nematic transition and results in the formation of a metastable
but very long-lived rigid cellular structure of very thin
interfaces, where the particles are densely packed, encapsulating
large volumes of nearly pure nematic liquid. We develop a
mean-field theoretical model describing the phase stability and
the mechanism of cellular solid formation.

\section{Sample preparation and experimental methods}   \label{expt}
In order to investigate the robustness of the process discovered
by Meeker et al. \cite{wilson}, we study two materials: 5CB and
MBBA, both archetypical thermotropic nematics \cite{lcbase}. The
case of $4'$-pentyl-$4$-cyanobiphenyl, abbreviated as 5CB,
obtained from Aldrich Chemicals Co., has already been examined in
\cite{wilson}. This pure material has a stable nematic phase below
$T_{\rm ni}=35.8$C, followed by a crystal phase at $T_{\rm
x}=22$C. No crystallisation was observed down to $5$C when the
colloid particles were added.

In contrast, the MBBA
($N$-$4$-methoxybenzylidene-$4'$-butylaniline, from Aldrich),
although one of the most common nematic materials in the
literature, has a practical complication of being susceptible to
hydrolysis. The $C=N$ bond in the rigid molecular section
dissociates in the presence of water. The reaction is reversible,
meaning that at a given temperature and humidity there is an
equilibrium balance of proper MBBA compound and its hydrolysis
products, which act as impurities and weaken the nematic order.
The effect is most noticeable in the depression of the phase
transition points: in the ``dry'' MBBA the nematic transition
point is at $T_{\rm ni}^{(o)} = 47$C \cite{lcbase}, followed by a
crystal phase at $T_{\rm x}= 22$C, while we have used MBBA in
hydrolysis equilibrium at ambient conditions, with $T_{\rm
ni}=37.3$C and no crystallisation down to $5$C. We employed this
as a model for a ``dirty'', weakly nematic liquid.

Theoretical arguments suggest that the best phase separation
conditions are achieved for small colloid particles (the ones with
$WR/K \ll 1$, but not too small so that the reversal of this
relation can occur near $T_{\rm ni}$), see section~\ref{theomod}
below. We use monodisperse PMMA spheres (polymethylmetacrylate,
bulk glass transition at $T_{\rm g}=105$C) of radius $R=150 \,
\hbox{nm}$, with a polydispersity of about 0.04 (as determined by
dynamic light scattering). Particles were covered with chemically
grafted poly-$12$-hydroxystearic acid (PHSA) chains (prepared by
Dr. A. Schofield in Edinburgh). On grafting, the short stabilising
chains adopt a conformation radially extending from the grafting
surface, making a ``hairy'' particle. In isotropic suspending
liquids, these particles behave like almost ideal hard spheres. In
addition, in the nematic suspending matrix, the grafted chains
provide a homeotropic (radial) director anchoring with a typical
strength $W \simeq 10^{-6} \hbox{J/m}^2$. With the typical Frank
constant $K \simeq 10^{-11} \hbox{J/m}$ this makes the
dimensionless colloid parameter to take the value $WR/K \sim
0.02$.

The preparation procedure described in \cite{wilson} was followed.
The liquid crystal was added to the dried particles at room
temperature, i.e. while in the nematic phase (one does not expect
good mixing in this situation). We then raise the temperature to
well above $T_{\rm ni}$ and subject the sample to the ultrasonic
excitation. In this way the particles were gradually dispersed in
the isotropic phase of 5CB or MBBA; the samples were stored at
$T\sim 45$C (above $T_{\rm ni}$) in a tumbling device to ensure
that the mixtures were homogeneous before any experiment was
started. The particle concentration $\Phi$ has been measured by
weight.  Although the theoretical arguments require a proper
(global average) volume fraction $\Phi=N \, v_R/V$, we consider
the two adequately close because the densities of PMMA and of
nematic liquid are not very different. Differential scanning
calorimetry (DSC, Perkin-Elmer Pyris 1) was used to identify the
phase behaviour of the resulting mixtures.

For this work the major advantage of the confocal optical
microscope is that it can produce a three-dimensional image of a
relatively thick and optically inhomogeneous sample. Ordinary
polarised microscopy, commonly used in studies of liquid crystals,
does not produce results here because the aggregated colloid
samples strongly scatter light -- the samples are opaque. The
employment of a pinhole to eliminate the out-of-focus light from
planes above and below the focal plane in a confocal microscope
produces a sharp picture of optical contrast in this plane. We use
a Laser Scanning Confocal Microscope (LSCM 510, by K. Zeiss) in
the reflection mode which does not require fluorescent labelling,
illuminated by the monochromatic $540$~nm laser, without
polarisers. In this mode, the images could be collected at a depth
of up to $100 \, \mu$m under the top sample surface.

\section{Experimental results}
\subsection{Calorimetry}
 The starting point of practically any theoretical description of
thermotropic nematic colloid phase behaviour, e.g. \cite{roux}, is
the assumption that the nematic transition temperature in the
homogeneously mixed colloid is a linearly decreasing function of
particle concentration $\Phi$. In fact, below we shall explicitly
calculate the slope of this linear dependence: $\sim T^*\left( 1-
\alpha \Phi \,\right)$, eq.~(\ref{shift}). Of course, such a
linear variation means that the particles are regarded as fully
independent, non-interacting -- which cannot be true for high
concentrations. Also, this estimate for $T_{\rm ni}(\Phi)$ is
based on a value reversal of the colloid parameter $WR/K$ near the
transition. This requires a subtle balance of material constants.
If $WR/K$ remains large deep in the nematic phase, the particles
will quickly aggregate under the action of strong attraction
forces, see \cite{poulin98,martin}. If $WR/K$ remains small even
near $T_{\rm ni}$ in spite of its tendency to grow as the nematic
order parameter $Q(T)$ decreases (cf. section~\ref{theomod}), the
particles will not have a strong elastic energy around them and
the system will behave more like a molecular mixture, a nematic
with impurities -- a well studied subject with a different phase
morphology \cite{mixtures}.
\begin{figure} 
\centerline{\resizebox{0.47\textwidth}{!}{
\includegraphics{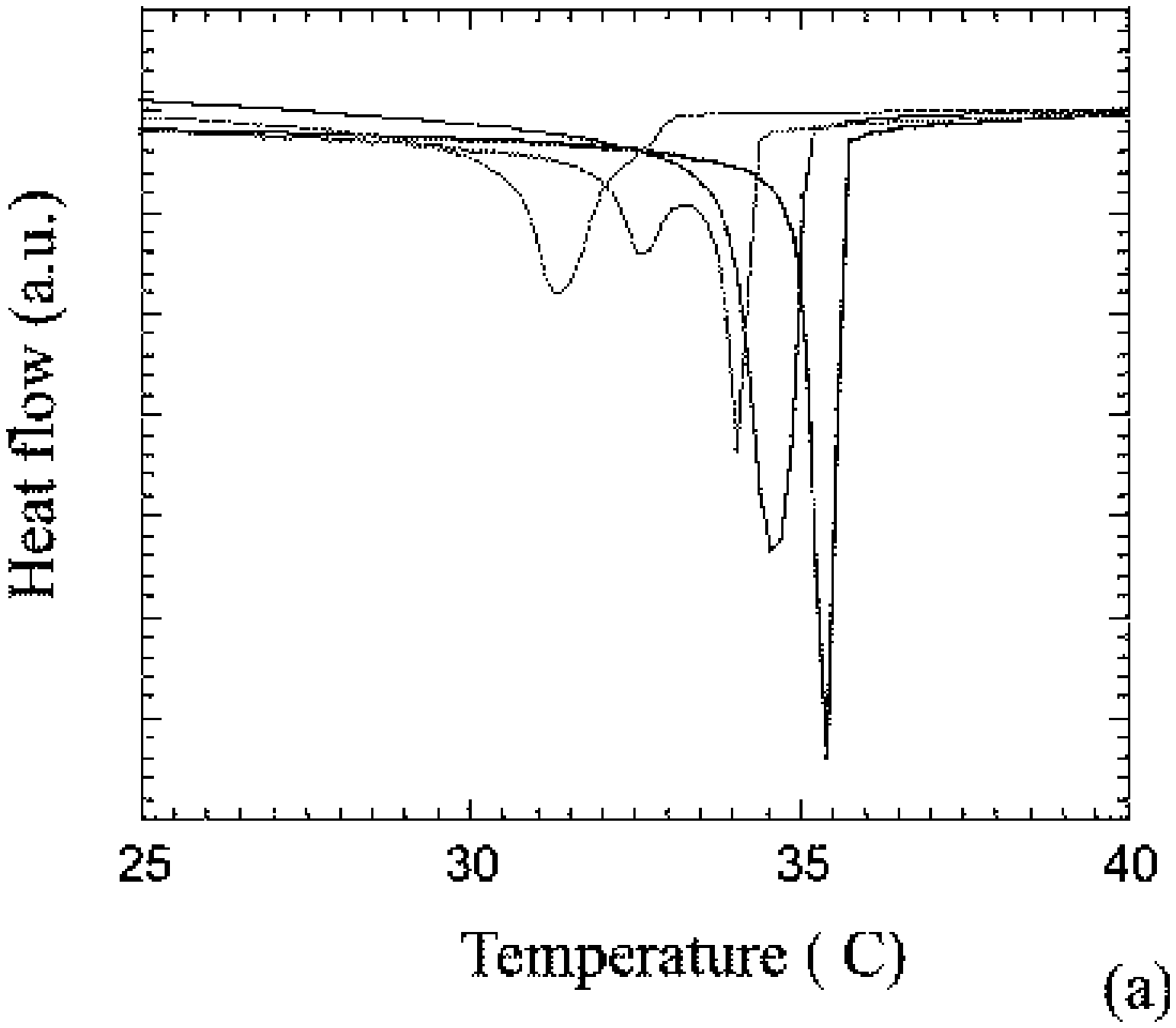} \hspace{0.5cm} \includegraphics{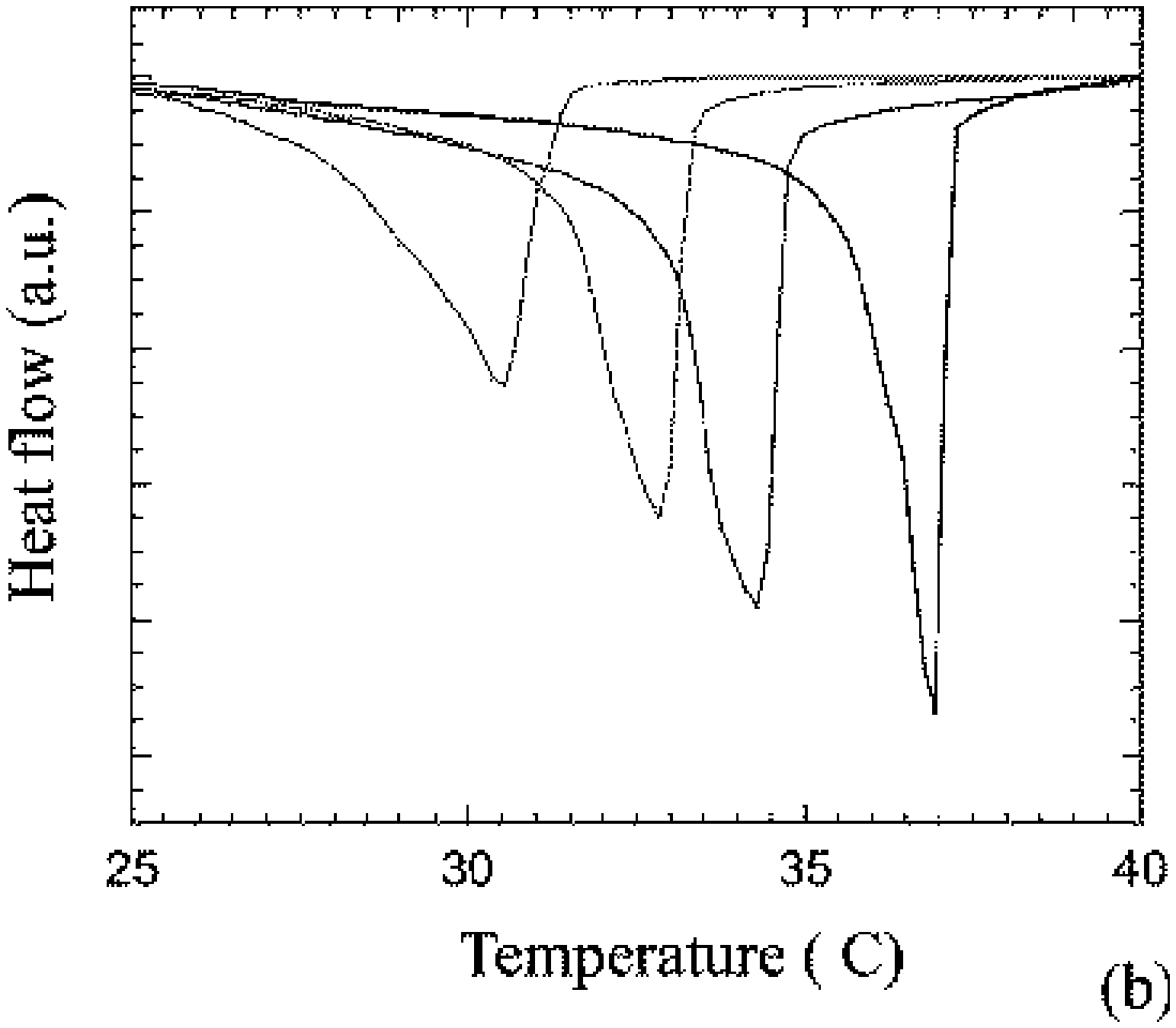}
}}
 \caption{Differential scanning calorimetry of nematic
transition on cooling a nematic colloid mixture from its
homogeneous isotropic state. The heat flow peaks are exothermic,
cooling rate $5^{\rm o}/\hbox{min}$.  (a) Data for 5CB at
concentrations $\Phi=0, \, 3\%, \, 7\% $ and $15\%$ (consecutive
curves with decreasing $T_{\rm ni}$). One should note both the
shift of the transition temperature and the broadening of the
peaks with increasing colloid concentration. (b) Data for MBBA at
concentrations $\Phi=0, \, 4.5\%, \, 8\%$ and $14.5\%$ (also
consecutive curves with the peak moving from right to left). }
\label{dsc}
\end{figure}

\begin{figure} 
\centerline{\resizebox{0.3\textwidth}{!}{
\includegraphics{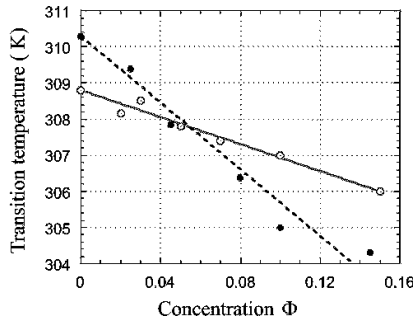}}}
 \caption{The variation of nematic transition temperature with
particle concentration (average volume fraction $\Phi$), obtained
from the DSC analysis. The data for $T_{\rm ni}(\Phi)$ for 5CB and
MBBA nematic colloids are fitted with the linear functions:
$308.8(1-0.06 \Phi)$ for 5CB - $\odot$ and solid line, and
$310.3(1-0.16 \Phi)$ for MBBA - $\bullet$ and dashed line. }
\label{tni}
\end{figure}

 In order to determine the coupling energy between the nematic
field and the individual colloid particle we study the
isotropic-nematic phase transition on cooling from a homogeneously
mixed state with a small particle volume fraction $\Phi$. The
differential calorimetry shows an exothermal latent heat peak at
this weakly first-order transition. Figure~\ref{dsc} shows the
evolution of this peak on increasing the particle concentration,
in 5CB and MBBA colloids. Two main features have to be noted.
First, the dependence of $T_{\rm ni}$ (determined by the peak
onset) on concentration shows a monotonic trend, fig.~\ref{tni}.
The linear fits appear to be quite good, with the slope consistent
with theoretical estimates of section~\ref{theomod}. Secondly, one
finds that the peaks become broader and of lower total area on
increasing the particle concentration. This is not unexpected, the
first order transition should become more and more diffuse with
added impurities. In fact, there is a theoretical view
\cite{aw,cardy} that such a transition should become continuous in
equilibrium. We did not address this question in detail, which
would have required significantly lower DSC cooling rates, at the
very least.

\subsection{Cellular structure}
\begin{figure} [hb]
\centerline{\resizebox{0.47\textwidth}{!}{
\includegraphics{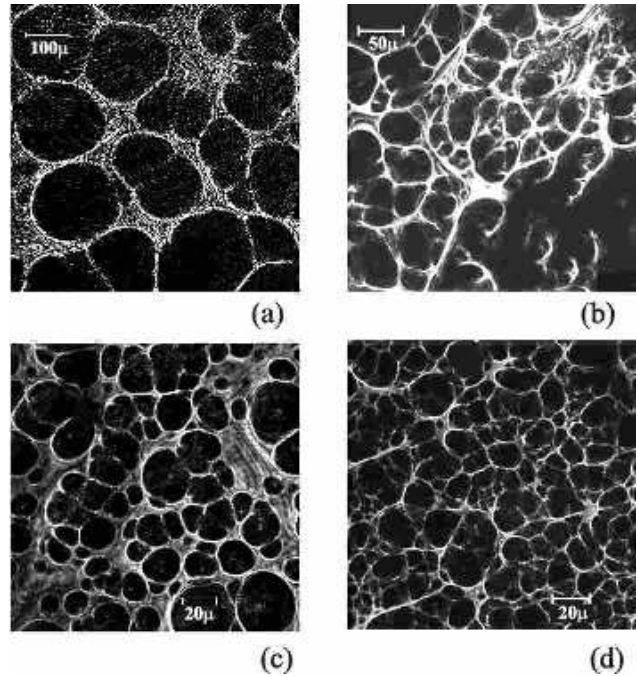} }}
 \caption{Reflection-mode confocal images of cellular structures.
The micrographs show a sample cross-section at the depth $\sim 50
\, \mu\hbox{m}$, for the 5CB-based colloid at room temperature and
particle concentrations $\Phi=3\%$ (a), $7\%$ (b), $10\%$ (c) and
$15\%$ (d). } \label{cells}
\end{figure}
Here we present the confocal images of horizontal cross-sections
of aggregated nematic colloids. The samples studied here were
prepared by simply depositing a portion of isotropic homogeneous
mixture (at $45$C) on a slide and allowing it to cool to the room
temperature, without top cover. The contrast mechanism between the
nearly pure nematic regions and the densely aggregated PMMA
particles is due to a different average refractive index in the
reflection mode of the microscope.

\begin{figure} 
\centerline{\resizebox{0.47\textwidth}{!}{
\includegraphics{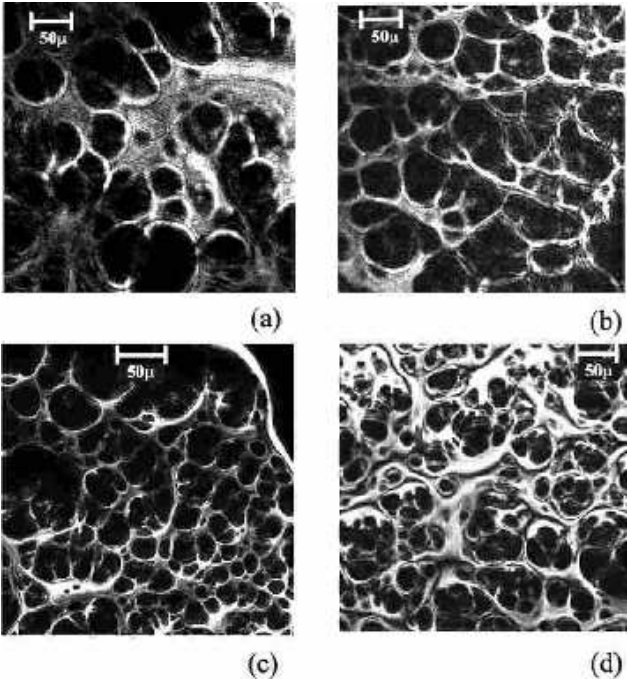} }}
 \caption{Reflection-mode confocal images of cellular structures.
The micrographs show a sample cross-section at the depth $\sim 50
\, \mu\hbox{m}$, for the MBBA-based colloid at room temperature
and particle concentrations $\Phi=2.7\%$ (a), $4.5\%$ (b), $10\%$
(c) and $14.5\%$ (d). } \label{cellsM}
\end{figure}

The cellular morphology of aggregated nematic colloids is apparent
in fig.~\ref{cells} for the 5CB system and in fig.~\ref{cellsM}
for the MBBA system. The relatively large cavities are separated
by thin interfaces providing an optical contrast. The proportion
of volume between these two fractions indicates that the cavities
are filled with the nematic liquid (the majority phase) and the
interfaces are made of PMMA particles.

The sizes of cells appear very irregular. One of the reasons is
that the thickness of confocal imaging plane is much smaller than
the cells and one obtains their different cross-sections. However,
it is also clear that cells are very polydisperse both in size and
in shape. One may argue that the way of sample preparation has
caused the polydispersity: the relatively slow cooling under
ambient conditions passes through the binodal regime of nucleation
and growth in the phase diagram, see later (fig.~\ref{fig7b}), and
would result in a variety of sizes of growing nematic nuclei. This
should be contrasted with a rapid cooling deep into the spinodal
decomposition regime, where a characteristic size is selected. We
shall see later that samples cooled in the rheometer (at a rate of
$30^{\rm o}/\hbox{min}$, with an additional small vibration aiding
interface packing) have a significantly more regular cellular
superstructure.

\begin{figure} 
\centerline{\resizebox{0.47\textwidth}{!}{
\includegraphics{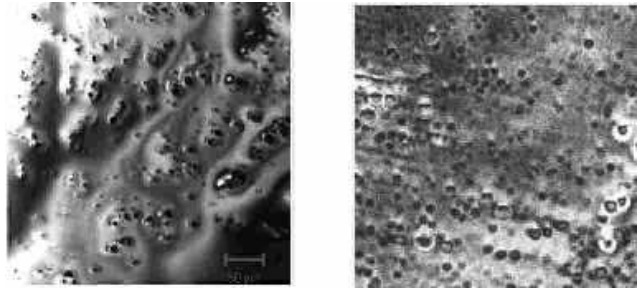}}   }
 \caption{Reflection-mode micrographs of free surface of
aggregated colloids in ambient open-air conditions. Left: 5CB
sample with $\Phi=10\%$, right: MBBA sample with  $\Phi=8\%$. }
\label{open}
\end{figure}

Another fact, evident in the confocal images, figs.~\ref{cells}
and \ref{cellsM}, is that the cell walls are perforated. This is
especially evident in the image (b) for the 7\% colloid, where
most of the walls are continuous in the cross-section, but some
are clearly interrupted. This observation brings our system into
the class of ``open-cell'' structures, cf. \cite{ashby}.
Figure~\ref{open} shows the microstructures of a free surface of
aggregated cellular nematic colloid prepared in the ambient
conditions. The difference with the internal cellular morphology
is striking. Nevertheless, in spite of a much more regular
surface, representing an effective cell wall meeting the outer
interface, the perforations of the interface are evident.

Image analysis, which should provide the statistical distribution
of cell sizes, is difficult for the confocal scans of different
cell cross-sections. Even if the cells were monodisperse, their
areas crossing the focal plane would appear broadly distributed.
Although there is a clear decrease in cell size with increasing
particle concentration in fig.~\ref{cells}, the distribution is
too irregular and asymmetric to allow accurate quantitative
conclusions.

\begin{figure} [ht]
\centerline{\resizebox{0.47\textwidth}{!}{
\includegraphics{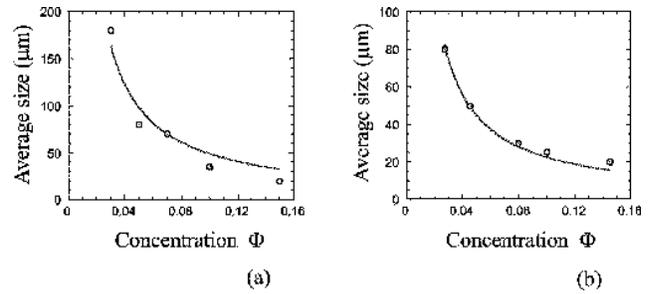}  }}
 \caption{Analysis of the average cell size $\lambda$
as function of initial particle concentration, at room
temperature, for PMMA particles of $R=150 \, \hbox{nm}$ dispersed
in (a) 5CB and (b) MBBA. Solid lines show the fits to $\lambda =
4.88/\Phi$ or $n \approx 32$ (5CB), and $\lambda = 2.22/\Phi$ or
$n \approx 15$ (MBBA). } \label{sizes}
\end{figure}

\section{Theoretical model}\label{theomod}
 First of all, let us re-iterate that the proposed regime
of phase separation as a mechanism of particle aggregation in the
nematic phase is taking place due to the smallness of the liquid
crystal colloid parameter $WR/K$. In the opposite case, the study
of which has been attempted several times over the years, the
topological defects around particles result in strong interaction
forces. If there are only few large particles in the system, they
often form ``strings'' -- chains of alternating spheres and
topological defects \cite{luben,poulin98}. When the particle
concentration is finite, the colloid undergoes fast aggregation
into 3-dimensional flocs. Big flocs with stronger relative
anchoring (planar on silica surface, as in \cite{martin}) fall
into the category of $WR/K\gg 1$. An intermediate stage of
evolution is the network of disclinations connecting the flocs and
having an effective modulus $G' \sim 0.01 \, \hbox{Pa}$. The
topological defect network would eventually reach the global
equilibrium by clearing the nematic volume altogether. (This final
state is the eventual fate of all other metastable modes of
aggregation, including our cellular solids).

 We now return to the case of weak anchoring, $WR/K \ll
1$, and consider the behaviour of a system of small particles near
the nematic transition, where the amplitude of the nematic order
parameter $Q(T)$ becomes small (the N-I transition is weakly first
order, so that the latent heat and the jump of $Q$ at the
transition are small). Both relevant elastic parameters, the Frank
constant $K$ and the anchoring energy $W$ vanish as $Q \rightarrow
0$, but with different rates: $K \sim \kappa Q^2$, while $W \sim w
Q$ in the first approximation. Therefore, in the immediate
vicinity of $T_{\rm ni}$ the ratio $WR/K \propto 1/Q$ should
become large and the topological regime of nematic director
distortions around even very small particles may prevail. This
would certainly be the case if a critical second-order transition
occurred at $T_{\rm ni}$. As it is, since $Q$ never reaches zero,
only a certain combination of material parameters would allow
$WR/K$ to change from small to large.\footnote{If $WR/K$ remains
small throughout the nematic phase, the elastic energy of
deformations, $\sim W^2R^3/K$ \cite{ukra}, has only a weak
dependence on the nematic order parameter $Q$ and results in long
aggregation times.} The elastic energy of these distortions is,
near the N-I transition, $\sim 10 KR = 10 \kappa R \, Q^2$. This
makes an addition to the Landau free energy density describing the
nematic transition, for local particle concentration $\phi$ it
takes the form $\sim 10 K R\, \phi / v_{\rm R}$, with $v_{\rm
R}=\frac{4}{3}\pi R^3$ the particle volume, giving
\begin{equation}
F_{\rm n} =  \frac{1}{2}A_{\rm o} \left[T-T_c(\phi) \right] Q^2 -
\frac{1}{3}B \, Q^3 + \frac{1}{4}C \, Q^4   \label{landau1}
\end{equation}
Here the shift in the critical point is due to the elastic energy
around the particles:
\begin{equation}
T_c(\phi)= T^*\left(1 -\frac{10\kappa R \, \phi}{ A_{\rm o}T^*
v_{\rm R} }\right)  \sim T^*\left( 1- \frac{\xi^2}{R^2} \, \phi
\,\right) \label{shift}
\end{equation}
where the second, very approximate, equation uses the fact that
the nematic correlation length $\xi$ is actually given by the
ratio of parameters $\kappa$ and $A_{\rm o}$, see Appendix. The
experimental data in fig.~\ref{tni} show that the slope of
$T_c(\Phi)$ is indeed noticeable. The experimental estimate
obtained from the linear fit is not wholly unreasonable, albeit
slightly larger than one could expect for our particles and a
typical correlation length $\xi \sim 10 \, \hbox{nm}$.

\subsection{Phase separation below $T_{\rm ni}$} \label{phasesec}
Immediately after the transition, the growing nematic nuclei start
expelling the particles in the drive to reduce its thermodynamic
free energy (\ref{landau1}). The initial average particle volume
fraction $\Phi$ splits into locally different valuesin phase
separating regions; we now use the notation $\phi$ for the local
concentration. The equilibrium nematic order parameter of the
mixed phase is
\begin{equation}
Q^*=\frac{B}{2C} \left( 1+ \sqrt{1+\frac{4(A_{\rm o} T^*)C}{B^2}
[1-\tau - \alpha \phi]} \right) \label{q}
\end{equation}
with the reduced temperature $\tau=T/T^*$ and the coefficient
$\alpha \sim 0.1$, using the data of fig.~\ref{tni}. Accordingly,
the upper limit of stability of homogeneously mixed nematic phase
is at the concentration
\begin{equation}
\phi_{\rm N} = \frac{1}{\alpha} \left( \frac{B^2}{4( A_{\rm o}T^*)
C} + 1- \tau \right) . \label{limphi}
\end{equation}
The equilibrium free energy density $F_{\rm n}(Q^*)$ then becomes
a function of local particle concentration $\phi$, see
eq.~(\ref{fullF}) below. The principal feature of this optimised
free energy density of a nematic mean field is the rapid increase
of $F_{\rm n}(\phi)$ with increasing particle concentration and
the non-convex form of this dependence. This feature is the
driving force for phase separation.

This energy penalty on having the particles uniformly dispersed in
the nematic matrix has to be added to the free energy of mixing.
The Carnaham-Starling excess free energy density of a hard-sphere
suspension \cite{fluid} is a very good approximation of the
equation of state:
 \begin{eqnarray}
F_{\rm p} &=& \frac{kT}{v_R} \left(\phi \, \ln \phi + \phi^2
\frac{4-3\phi}{(1-\phi)^2}
 \right) , \ \ \ \phi<\phi^*, \ {\rm liquid} \label{fluid} \\
F_{\rm p} & =& \frac{kT}{v_R} \left( 1.79 \phi + 3 \phi \, \ln
 \frac{\phi}{1-\phi/\phi_c} , \right) , \ \ \phi >\phi^*, \
 {\rm solid } \nonumber
\end{eqnarray}
where $v_R$ is the particle volume and the two expressions are
matched by an adjustable parameter $u\approx 1.7929$ at a
concentration $\phi^* \approx 0.52$, with $\phi_c \approx 0.64$
the random close-packing fraction. The full free energy of nematic
colloid, $F_{\rm n} + F_{\rm p}$, then becomes, in dimensionless
form,
\begin{eqnarray}
&& \frac{v_R}{kT} F = -\frac{a_2}{\tau} (1-\phi)\bigg[ (1-\tau -
\alpha \phi)^2 \label{fullF}  \\ &+& \frac{b_2^2}{6} \left(1+
\frac{6}{b_2} (1-\tau -\alpha \phi) + \big[ 1+ \frac{4}{b_2}
(1-\tau -\alpha \phi) \big]^{3/2} \right) \bigg] \nonumber \\ &&
\qquad \phi \, \ln \phi + \phi^2 \frac{4-3\phi}{(1-\phi)^2} \ ,
\nonumber
 \end{eqnarray}
at small concentrations ($\phi < \phi^*$), where the dimensionless
parameters $a_2 \gg 1$ and $b_2 \ll 1$ are estimated in Appendix,
eq.~(\ref{par2}). The factor $(1-\phi)$ in front of the nematic
mean field energy $F_{\rm n}(Q^*)$ measures the local proportion
of the solvent in the colloid mixture.

\begin{figure}
\centerline{\resizebox{0.35\textwidth}{!}{
\includegraphics{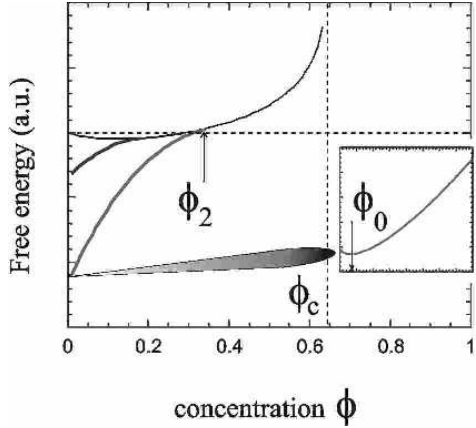}    }}
 \caption{Plots of free energy
density (\ref{fullF}) for decreasing temperature: $T>T^*\approx
309$ ($Q=0$, thin solid line, eq.~\ref{fluid}), $T=305$ and
$300^{\rm o}$K.  The nematic energy $F_{\rm n}$ in (\ref{fullF})
has been scaled by a factor $\sim 10^{-3}$ to let it fit on the
same graph. At concentration $\phi_2(T)$ the nematic nematic
transition takes place and the isotropic state with $Q=0$ takes
over. The inset shows the fine structure of the plots very near
$\phi=0$ with a well defined minimum at $\phi_0$. } \label{fig7a}
\end{figure}

 This is the simplest possible model of nematic colloid phase
ordering, essentially the same as used in \cite{roux}, with minor
deviations in the interpretation of nematic order-composition
coupling and material constants. This interpretation, however,
determines the results and predictions. All parameters entering
the free energy $F(\phi)$ will be determined from experiment, thus
providing the phase diagram in the temperature-composition plane
with no adjustable free parameters.

At low initial average particle volume fraction, the structure of
eq.~(\ref{fullF}) is reminiscent of the Flory-Huggins model for
demixing. The nematic mean-field energy plays the role of the
effective $\chi$-parameter potential term in that model, providing
non-convex variation in $F(\phi)$. Figure~\ref{fig7a} shows a
series of plots of eq.~(\ref{fullF}) for the set of Landau
parameters corresponding to 5CB nematic liquid crystal,
eqs.~(\ref{params}) and (\ref{par2}) in Appendix. There is always
a well-defined minimum at very small concentrations, $$\phi_0
\simeq \exp \left[- a_2 \, \frac{(1-\tau) (1-\tau+2\alpha)}{\tau}
\right],$$ due to the entropy contribution $(\phi \, \ln \phi)$ in
eq.~(\ref{fluid}). The usual common-tangent construction for the
$F(\phi)$ connects two branches of the free energy: one
non-convex, at very low $\phi$, in an almost pure liquid crystal,
and the other fully convex, in the high-$\phi$ phase where the
nematic order is unstable $(Q=0)$ and the particles are densely
compacted.
\begin{figure} 
\centerline{\resizebox{0.35\textwidth}{!}{
\includegraphics{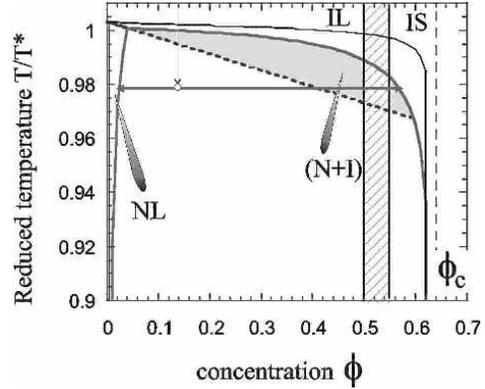}
}}
 \caption{The phase
diagram of a nematic colloid, for the choice of parameters
discussed in Appendix (thin solid lines) and for an artificially
small nematic mean field, scaled by $\sim 10^{-3}$ to make the
phase boundaries apparent (solid binodal lines). In the isotropic
phase, the hard-sphere colloid is in the liquid state (IL) below
the coexistence boundary $\phi \simeq 0.5$ and is in the solid
state (IS) at high $\phi \rightarrow \phi_c$. See text for
explanations of other phase boundaries. } \label{fig7b}
\end{figure}

The schematic phase diagram of the nematic colloid is shown in
fig.~\ref{fig7b} in coordinates of reduced temperature $\tau$ and
concentration $\phi$. At high temperature the colloid behaves as a
standard athermal hard-sphere system, which crystallises above
$\phi \sim 0.5$ and reaches the maximum random close-packing level
at $\phi_c \sim 0.64$. At an extremely low particle concentration,
corresponding to the expanded insert in fig.~\ref{fig7a}, there is
a region of homogeneous nematic mixture, labelled NL in the phase
diagram. This region is artificially expanded in fig.~\ref{fig7b},
the real numbers make it invisible in the plot. A very good
analytical approximation for this phase boundary is
 \begin{equation}
\phi_1(\tau) \simeq \frac{\tau}{1-\tau+\frac{1}{2}\alpha} \,
\left(\frac{1}{ a_2 \, \alpha}\right) \sim 10^{-6} \tau ...
 \end{equation}

The spinodal on the high-$\phi$ branch is determined by the point
of contact between the nematic (concave, locally unstable) and the
isotropic (fully convex) parts of the free energy, which is marked
as $\phi_2(\tau)$ in fig.~\ref{fig7a}. Its analytical estimate,
$$\phi_2=\frac{1}{\alpha} \left( \frac{2}{9}b_2 + 1- \tau
\right),$$ is practically indistinguishable from the stability
limit $\phi_{\rm N}$ in eq.~(\ref{limphi}). This expresses the
known fact about the ``weakness'' of first order nematic
transition: the width of nematic coexistence is very small in
ordinary thermotropic liquid crystals. Below the spinodal line
$\phi_2(\tau)$ the homogeneous particle mixture is unstable. Above
this line, the shaded region in fig.~\ref{fig7b} shows the zone of
coexistence between the low-$\phi$ (nematic) and the high-$\phi$
(isotropic) phases. The upper boundary of this region, the
high-concentration binodal line is calculated numerically from the
common-tangent condition for $F(\phi)$ in fig.~\ref{fig7a}, for
the set of parameters given by eq.~(\ref{par2}) (thin solid line)
and the artificially scaled down nematic energy (bold solid line)
to expose the critical point, where this binodal merges with the
low-$\phi$ phase boundary $\phi_1(\tau)$.

Note that the high-concentration phase is in the state of
colloidal solid. This is a consequence of very large relative
strength of nematic mean field, expressed by the large
dimensionless parameter in eq.~(\ref{par2}). One may say that the
effective pressure from the nematic liquid to expel the particles
and compact them in high-$\phi$ regions is rather high.

A word of caution has to be sounded at this point, regarding the
continuing usage of the linear law $T_c(\phi)$ at relatively high
particle concentrations. This approximation is not likely to be
valid in the regime of high interparticle interactions and,
therefore, the spinodal $\phi_2(\tau)$ would deflect down from the
straight line in fig.~\ref{fig7b}. We, however, are mostly
interested in small particle concentrations where the
approximation of independent particles would hold.

\subsection{Cellular structure} \label{cellsec}
 We thus envisage the following
mechanism of phase evolution of the nematic colloid on its cooling
from the isotropic homogeneously mixed phase:
\begin{itemize}
\item  After quenching the homogeneous colloid
suspension below the nematic transition point, the system phase
separates into growing regions of pure nematic liquid, from which
the particles are expelled into the boundaries.
\item On these interfaces of growing nematic nuclei, the
concentration of particles is so high that the colloid solidifies
and the remaining mesogenic liquid is in the isotropic state.
\end{itemize}
Clearly, the fine balance of kinetic effects is essential: the
phase separation should have a higher rate to develop within
growing nematic nuclei. The driving force for this process is the
thermodynamic free energy gain in having the high nematic order in
a clean system at a given $T<T_{\rm ni}$, in comparison with the
uniformly dispersed nematic colloid, which is expressed by
non-convexity of plots in fig.~\ref{fig7b}.
\begin{figure} 
\resizebox{0.47\textwidth}{!}{
\includegraphics{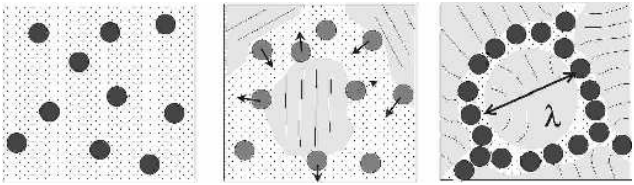} }
\caption{Model of particle phase separation showing the
homogeneous mixing in isotropic phase, the growing nematic nuclei
expelling the particles and the final state, where the particles
are densely packed on thin interfaces under the effective pressure
from the nematic mean field in the encapsulated domains of size
$\lambda$. } \label{modelpic}
\end{figure}

Eventually, when the clean nematic regions grow to come into the
near-contact with each other, the particles are only allowed to
pack on narrow interfaces. The sample transforms into a cellular
superstructure with rather thin densely-packed walls encapsulating
nematic volumes of characteristic size $\lambda$. This cell size
is determined by the initial average colloid concentration $\Phi$
and the thickness of interfaces. Let us define this thickness as
$d\equiv nR$, several times the particle radius (the smallest
possible is $n\sim 2$ in fig.\ref{modelpic}). Then the total area
of such interface in the sample volume $V$ is ${\cal A}=\Phi
V/(nR)$. This leads to the order of magnitude estimate for the
mesh size
\begin{equation}
\lambda = (V/{\cal A}) \sim \frac{nR}{\Phi}  . \label{mesh}
\end{equation}
This gives $\lambda \sim 6 \, \mu\hbox{m}$ for this perhaps
unrealistic case of $n=2$, and $R=150 \, \hbox{nm}$ and
$\Phi=0.05$. The analysis of structures in the next section
obtains $n \sim 20$, see fig.~\ref{sizes}, i.e. interfaces of
$\sim 10$ particles thick and $\lambda \sim 60\, \mu\hbox{m}$ for
$\Phi=0.05$.
\begin{figure} 
\centerline{\resizebox{0.2\textwidth}{!}{
\includegraphics{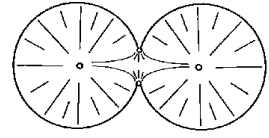} }}
\caption{An illustration of topological barrier for the breaking
of interface between two cells, sufficiently big to sustain the
topological charge, $\lambda \gg K/W$, see text.} \label{drops}
\end{figure}

The cellular solid with thin high-tension interfaces is not the
globally equilibrium state. It is obvious that the free energy can
be further lowered by aggregating all particles into a
3-dimensional densely packed group, leaving the whole volume for
the pure nematic liquid phase. However, the metastable cellular
structure may become frozen, trapped by high energy barriers for
disrupting the formed interfaces. Several mechanisms contribute to
creating such barriers. Possibly the largest contribution arises
from the nematic topological argument similar to that in liquid
crystal emulsions \cite{emuls95}. Consider the nematic alignment
in one cell, regarding the interface as a continuous wall imposing
a homeotropic director anchoring energy $W$. When the mesh size
$\lambda$ increases so that the familiar dimensionless parameter
$(W\lambda /K) \gg 1$, the nematic liquid crystal within this cell
has to possess the topological charge $N=1$. When one attempts to
break through the cell interface, two topologically charged
volumes come into contact and their coalescence must comply with
the law of this charge conservation. Therefore, in the moment of
formation of a channel between the cells, a new topological defect
of the charge $N=-1$ must be created \cite{emuls95}. Later this
defect may travel towards one of existing monopoles and annihilate
it to minimise the elastic energy of the final joint volume.
However, this cannot happen instantly and the formation of each
new topological defect costs an elastic energy $\sim K \lambda$.
This energy barrier can be very high: $\sim 10^{-16} \hbox{J}$,
compared with $k_{\rm B}T \sim 4 \times 10^{-21} \hbox{J}$, and
the system may remain trapped in the metastable state of random
cellular structure. One then expects to find the effects of ageing
and time translation invariance breaking, characteristic of weakly
non-ergodic glassy dynamics and rheology \cite{bouchaud,sollich}.

\section{Conclusions}  \label{concl}

We have reported the results of a structural study of nematic
colloids based on a classical thermotropic liquid crystals matrix
with small monodisperse polymer particles suspended in it
\cite{wilson}. The hydrophobic sterical stabilisation ensures the
radial anchoring of the director on particle surface. At
relatively small concentrations of particles, we observe good
mixing above the clearing point of isotropic-nematic phase
transition $T_{\rm ni}$ and a rapid aggregation of the homogeneous
mixture into a rigid gel-like solid, completely opaque optically
at and below $T_{\rm ni}$. The properties of the phase ordering
and the morphology of the phase-separated aggregates were the
primary focus of this study.

On cooling from the homogeneous isotropic mixture, we observe a
decrease in the transition temperature $T_{\rm ni}$ as a function
of average colloid concentration $\Phi$, which follows a
reasonably linear law in the region of small concentrations
studied here. This is an expected and frequently observed
phenomenon, accompanied by a noticeable diffusion of the weak
first-order phase transition, which is otherwise sharp in a pure
nematic liquid crystal.

Below $T_{\rm ni}$ the colloids undergo phase separation with the
resulting structure best described as an open cellular solid, with
the particles densely packed in thin walls and the cavities filled
with a pure liquid crystal. The characteristic cell size is of the
order of $10-100~\mu$m and appears to follow the
inverse-proportionality law $\lambda \sim 1/\Phi$, also confirmed
by a simple theoretical estimate. Since the nematic director is
anchored on cell walls, the resulting randomly quenched
birefringent texture strongly scatters light, giving the material
its opaque appearance. The remarkable mechanical rigidity of the
resulting cellular solids, stemming from the high effective
surface tension of interfaces, is the subject of the companion
paper \cite{no2}.

Colloid suspensions in nematic liquid crystal matrices,
thermotropic and lyotropic, have been studied before. Particle
aggregation is an unavoidable result in all cases, since the
system below $T_{\rm ni}$ tends to minimise the elastic energy of
director distortions around individual particles. However, the
cellular solid morphology, first observed in \cite{wilson}, is
new. We believe that two main factors contribute to the reason why
we obtain such an effect. The particles have to be small and have
sufficiently low anchoring energy to ensure that the colloid
parameter $WR/K$ is small just below $T_{\rm ni}$. On the other
hand, our theoretical model suggests that this parameter has to
become substantial again, to provide a required thermodynamic
force for phase separation. This rather restricts a range of
particle sizes and surface treatments that allow for such a
compromise to occur. Secondly,  we find that the rate of cooling
through the first-order nematic transition has an important
effect. The cellular structure is much more regular and robust
when the homogeneous isotropic colloid is quenched rapidly deep
below $T_{\rm ni}$. The theoretical phase diagram suggests that in
order to achieve a good selection of size one needs to avoid a
binodal region just below $T_{\rm ni}$. In many cases of slow-rate
cooling and a colloid material parameters making the binodal gap
wider (such as in our example of MBBA system), the mixture would
separate in a different fashion -- perhaps totally avoiding the
cellular solid regime.

Although the details of aggregation mechanism and forces that hold
particles together at interfaces are somewhat unclear, the
suggested theoretical ``toy model'' offers a possible explanation
for the formation of cellular structure. Still, many questions
remain open. Such an unusual phase behaviour and remarkable
rheological properties of the liquid crystal colloid suspensions
require further detailed study, both theoretical and experimental.
\\

\noindent We appreciate valuable discussions with M.E. Cates, S.M.
Clarke, P.D. Olmsted and M. Warner. The help of I. Hopkinson with
the confocal microscope is gratefully acknowledged. This research
has been supported by the UK EPSRC.

\appendix
\section*{Appendix:\\ Parameters of Landau free energy}
It is useful to find the values of the three phenomenological
parameters describing the transition, coefficients $A_{\rm o},\,
B$ and $C$ in eq.~(\ref{landau1}), although different materials
will have these parameters slightly different. It is nevertheless
instructive to examine the characteristic orders of magnitude. In
order to determine three parameters one needs three independent
measurements. We are fortunate that there are, in fact, four
available:

1) The jump of order parameter at the weak first-order transition
is $\Delta Q_{\rm ni}=2B/3C$. There might be some error in its
determination, which should depend (among other factors) on the
rate of cooling through the transition. However, because $Q$ only
varies between $0$ and $1$ and because a number of molecular
theories predict this jump explicitly, one can take qualitatively
$Q_{\rm ni}\simeq 0.4$, giving $B \approx 0.6C$.

2) The second measurement can be the interval between the
transition temperature and the critical point $T^*$ (the latter
may be determined by extrapolating the inverse susceptibility,
$\chi^{-1} \sim |T-T^*|$). In usual thermotropic nematic liquid
crystals this interval is rather small, $T_{\rm ni}-T^* =2B^2/9
A_{\rm o}C \sim 1^{\rm o}$. This gives $B\approx 7.5 A_{\rm o}
\cdot 1^{\rm o}$K. (Note that the dimensionality of $B$ and $C$ is
energy density, while $A_{\rm o}$ has the dimensions of
$\hbox{J/m}^3{}^{\rm o}\hbox{K}$).

3) The latent heat of the first-order phase transition is $\Delta
H =T_{\rm ni} (2 A_{\rm o}B^2/9C^2)$, per unit volume. It is
typically obtained from calorimetry by integrating the
characteristic peak. Again, a large uncertainty may accompany such
a measurement because at any non-zero cooling rate the conditions
are not exactly equilibrium. However, keeping up the qualitative
approach, the value of transition enthalpy per unit mass is of the
order $ \sim 1 \, \hbox{J/g}$ for 5CB at a reasonably slow cooling
rate. Taking the density $\rho \sim 1 \, \hbox{g/cm}^3$ and the
$T_{\rm ni} \sim 310^{\rm o}$K, we obtain
\begin{eqnarray}
A_{\rm o} & \sim & 6 \times 10^3 \hbox{J/m}^3{}^{\rm o}\hbox{K}
\label{params} \\ {\rm Then} \ \ B & \sim &  5 \times 10^4 \ \
{\rm and} \ \ C\sim 1.2 \times 10^5 \hbox{J/m}^3.  \nonumber
\end{eqnarray}

4) It is important that the three measurements above give
estimates that agree with the fourth way of accessing $A_{\rm o}$.
The nematic correlation length $\xi$ may be determined by the
ratio of the bare Frank elastic constant $\kappa=K/Q^2$ to the
thermodynamic energy density: $\xi^2 = \kappa/A_{\rm o} \Delta T$.
There are many ways of confirming that the characteristic
magnitude of $\xi$ is $ \sim 10 \, \hbox{nm}$. For usual values
$\kappa \sim 10^{-11} \hbox{J/m}$ and $\Delta T\sim 10^{\rm o}$ we
obtain $A_{\rm o} \sim 10^4 \hbox{J/m}^3{}^{\rm o}\hbox{K}$, in a
very reasonable agreement with the previous estimate.

Finally, we may estimate the dimensionless factors entering the
final expression for the free energy (\ref{fullF}). At $T^*
\approx 309^{\rm o}$K ($kT^* \sim 4.7 \times 10^{-21} \hbox{J}$)
and particle volume $v_R \sim 1.4 \times 10^{-20} \hbox{m}^3$ we
obtain
\begin{equation}
a_2=\frac{A_{\rm o}^2 {T^*}^2 v_R}{4C \, kT^*} \sim 2\times 10^{7}
\ \ \  {\rm and} \ \ b_2=\frac{B^2}{A_{\rm o} T^* C} \sim 10^{-2}
. \label{par2}
\end{equation}

\end{document}